\documentstyle[emulateapj]{article}

\submitted{Accepted by the Astrophysical Journal}

\def\gsim{\;\rlap{\lower 2.5pt
 \hbox{$\sim$}}\raise 1.5pt\hbox{$>$}\;}
\def\lsim{\;\rlap{\lower 2.5pt
   \hbox{$\sim$}}\raise 1.5pt\hbox{$<$}\;}
\def\msun{{\rm\,M_\odot}}

\def\mic{{\,\mu{\rm m}}}

\def\K{{\rm\,K}}
\def\spose#1{\hbox to 0pt{#1\hss}}
\def\lta{\mathrel{\spose{\lower 3pt\hbox{$\mathchar''218$}}
     \raise 2.0pt\hbox{$\mathchar''13C$}}}
\def\gta{\mathrel{\spose{\lower 3pt\hbox{$\mathchar''218$}}
     \raise 2.0pt\hbox{$\mathchar''13E$}}}
\def\iun{{\rm\,nW\,m^{-2}\,sr^{-1}}}

\lefthead{AGUIRRE \& HAIMAN}
\righthead{INTERGALACTIC DUST AND THE FIRB}

\begin{document}\begin{flushright}
{\footnotesize
FERMILAB-Pub-99/195-A}
\end{flushright}
\nopagebreak
\vspace{-\baselineskip}

\title{Cosmological Constant or Intergalactic Dust? \\ Constraints
from the Cosmic Far Infrared Background}
\author{Anthony Aguirre}
\affil{Department of Astronomy, Harvard University\\
60 Garden Street, Cambridge, MA 02138, USA\\
email: aaguirre@cfa.harvard.edu}
\and
\author{Zoltan Haiman}
\affil{NASA/Fermilab Astrophysics Center\\ Fermi National Accelerator
Laboratory, Batavia, IL 60510, USA\\ email: zoltan@fnal.gov}

\begin{abstract}
  
  Recent observations of Type Ia SNe at redshifts $0<z<1$ reveal a
  progressive dimming which has been interpreted as evidence for a
  cosmological constant of $\Omega_\Lambda\sim 0.7$.  An alternative
  explanation of the SN results is an open universe with
  $\Omega_\Lambda=0$ and the presence of $\gsim 0.1\mic$ dust grains
  with a mass density of $\Omega_{\rm dust}\sim ({\rm few}) \times
  10^{-5}$ in the intergalactic (IG) medium.  The same dust that dims
  the SNe absorbs the cosmic UV/optical background radiation around
  $\sim 1\mic$, and re-emits it at far infrared (FIR) wavelengths.
  Here we compare the FIR emission from IG dust with observations of
  the cosmic microwave (CMB) and cosmic far infrared backgrounds
  (FIRB) by the DIRBE/FIRAS instruments.  We find that the emission
  would not lead to measurable distortion to the CMB, but would
  represent a substantial fraction ($\gsim75\%$) of the measured value
  of the FIRB in the $300-1000\mic$ range.  This contribution would be
  marginally consistent with the present unresolved fraction of the
  observed FIRB in an open universe.  However, we find that IG dust
  probably could not reconcile the standard $\Omega=1$ CDM model with
  the SN observations, even if the necessary quantity of dust existed.
  Future observations able to reliably resolve the FIRB to a flux
  limit of $\sim 0.5\,$mJy, along with a more precise measure of the
  coarse-grained FIRB, will provide a definitive test of the IG dust
  hypothesis in all cosmologies.

\end{abstract}
\keywords{cosmology: theory -- cosmology: observation -- cosmology: Far
Infrared Background -- cosmic microwave background -- galaxies: formation
-- galaxies: evolution}

\section{Introduction}

One of the most remarkable results in recent observational cosmology
is the detection of type Ia SNe at cosmological distances.  Comparing
the Hubble diagram of these $\sim50$ SNe with the predictions of
classical cosmological models strongly favors models with acceleration
in the cosmic expansion, presumably due to a significant cosmological
constant (Perlmutter et al.  1999; Riess et al. 1998).  Given the
far-reaching implications of this result, it is important to assess
the possible systematic effects that could mimic the behavior of cosmic
acceleration, and thus allow cosmologies without a cosmological
constant.

Intergalactic (IG) dust could provide such a systematic effect.  The
amount of dust required to fully account for the systematic dimming of
SNe at high redshift in an open universe is $\Omega_{\rm dust}\sim
({\rm few}) \times 10^{-5}$ (Aguirre 1999a; Aguirre 1999b [A99]). If
this quantity of dust had the same composition as interstellar dust in
the Milky Way, it would cause a significant reddening in the SN
spectra, which is not observed.  However, A99 showed that the removal
of small dust grains during their ejection from galaxies into the IG
medium could bias the composite dust cross section to cause little
reddening for a given amount of extinction. A99 presented specific
models for the evolution of the IG dust that are consistent with the
observed lack of reddening, but provide the amount of extinction to
fully account for the SNe results in an open universe.

In the present paper, we study the observational consequences of the
IG dust hypothesis in far infrared wavelength bands.  As is well known
from previous studies, IG dust would efficiently absorb the UV/optical
flux of galaxies and quasars, and re-radiate this energy in the far
infrared (e.g. Wright 1981; Bond, Carr \& Hogan 1991; Loeb \& Haiman
1997).  Although studied in great detail in these works, this effect
is at present of renewed interest, owing to recent direct estimates of
the global average comoving star formation rate (SFR) in the universe
(e.g. Madau 1999 [M99]), and determinations of the extragalactic
UV/optical background (UVB, Bernstein 1997), as well as of the cosmic
far infrared background (FIRB; see Fixsen et al. 1998; Puget et al
1996).  We thus have, for the first time, a measurement of both the UV
reservoir available for dust absorption, as well as the background
flux at the wavelengths where this energy would be emitted by the
dust.

In this paper, we compute contributions to the FIRB and CMB arising
from IG dust, and compare it to the measured values. Our aim is to
assess whether the amount and type of dust that would explain the SNe
results in an $\Omega_\Lambda=0$ universe is compatible with these new
observations.

At the present time, the source of the observed FIRB remains unclear,
although rapid progress is being made using deep observations with
SCUBA.  Barger et. al (1999a) and Hughes et al. (1998) have detected a
population of discrete IR sources whose cumulative flux down in flux
to $\sim 2\,$mJy accounts for at least 20-30\% of the FIRB at
850$\,\mu$m.  The gravitationally lensed sample of Blain et al.
(1999a,b) goes further down to $\sim 1\,$mJy, but is more uncertain
because of the lensing model and small number of detections.  It it
therefore currently uncertain whether the counts continue to
sufficiently faint flux to account for all of the FIRB.

Furthermore, we lack information on the spectra of
these sources at wavelengths other than 850$\,\mu$m.  Although the
full FIRB can be explained theoretically in semi-analytic galaxy
formation models (Guiderdoni et al. 1998), this requires the somewhat
ad-hoc postulate that the ratio of the number of ULIGs (a population
of ``ultraluminous infrared galaxies'') to that of normal optical
galaxies increases rapidly with redshift. In summary, according to our
present knowledge, a fraction 70-80\% of the FIRB could still be
contributed by diffuse emission from IG dust -- but this may change
very quickly.

As in the case of the FIRB, the nature of the observed UVB is unclear.
The directly observed values at 0.55 and 0.8$\,\mu$m are 20$\pm10\iun$
(Bernstein 1997).  By comparison, a recent compilation of ground based
galaxy counts, and a survey of galaxies in the Hubble Deep Field (HDF)
has yielded an estimate for the evolution of the global star formation
rate in the universe between $0<z\lsim 5$.  The UVB that follows from
a census of these galaxies is around $\sim 12\iun$
(Madau 1999 [M99]; Pozzetti et al.  1998), a factor of $\sim2$ smaller
than the observed value.  A natural explanation for the difference
would be if the remaining $\sim50\%$ of the UVB were contributed by
faint, undetected galaxies; however this appears to violate limits
from the fluctuations in the UVB measured in the HDF (Vogeley 1997).
Despite these uncertainties, the approximate amplitude of the UVB
allows us to compute useful estimates of the contribution of IG dust
to the FIRB.

The rest of this paper is organized as follows.  In \S~2 we summarize
our model for the amount and type of IG dust; in \S~3 we describe our
assumptions concerning the redshift evolution of both the UV emission
and dust production, and outline our calculation methods; and in
\S~4 we discuss limits from existing FIRAS observations of the CMB.
Section \S~5 contains our main results on the contribution to the FIRB
from IG dust, as well as hopes for future tests of the dust hypothesis
for the SN results; and, in \S~6 we summarize our conclusions.  Unless
otherwise stated, in this paper we adopt an open cosmology with total
matter density $\Omega=0.2$ and Hubble constant $h_{50}=1$.

\section{IG dust model}

The model adopted here for IG dust is discussed at length in A99; here we
provide only a brief summary.  The model is based on the following method of
estimating the total IG dust density $\Omega_{\rm dust}(z)$: we first estimate
the total metal density $\Omega_Z(z)$, then multiply this figure by the
fraction $F_I$ of these metals which lie outside galaxies, and then by the
fraction $d_m$ of IG metals is contained in IG dust.

As shown in A99, both a direct integration of the SFR (with an
assumption of the metals produced for each star formed) and fossil
evidence from clusters indicate that $\Omega_Z(z \la 0.5) \approx
(2.5-5) \times 10^{-4}$ (adjusted from A99 for $h_{50}=1$). These
estimates follow from conservative assumptions: that stars in cluster
galaxies have the same IMF as in field galaxies, that there is not a
dominant population of unobserved galaxies, etc.; see A99.

Measurements of the metallicity of intracluster gas indicate that a fraction
$F_I \approx 75\%$ of a typical cluster's metals lie in the intracluster medium
(e.g. Renzini 1997), presumably removed from the galaxies by some combination
of winds, dust expulsion, ram-pressure stripping, and tidal disruption/merging
of galaxies.  While ram-pressure stripping and mergers may be more effective at
removing metals from cluster galaxies, it is nevertheless likely that the
figure $F_I \approx 75\%$ applies to the field galaxies, since a much smaller
value, together with the estimate of $\Omega_Z \ga 10^{-4}$, would imply that
field galaxies have mean metallicity several times solar -- contrary to
observation.

To estimate the fraction $d_m$ of IG metals in dust, we assume a value
of $\sim 0.5$ for metals leaving the galaxy, as applies to the
interstellar medium of both typical galaxies and perhaps even to
damped Ly-$\alpha$ systems (Pei, Fall \& Hauser 1998).  Some fraction
$(1-f)$ by mass of this dust must be destroyed either during ejection
or in the IG medium, giving $d_m \approx 0.5f.$ Combining these
figures, we estimate an IG dust density at $z \la 0.5$ of
\begin{equation}
\Omega_{\rm dust} \approx (9.4-18.8)f \times 10^{-5}.
\end{equation}
To fix both $f$ and the dust properties, we shall adopt the two-component
Draine \& Lee (1984) [DL] dust model\footnote{The effects are, of course,
somewhat dependent upon the grain model; see A99 for discussion.} of
silicate and graphite spheres with size distribution $N(a)da \propto a^{-3.5},\
0.005\mic \le a \le 0.25\mic$ (Mathis, Rumpl \& Nordsieck 1977 [MRN]) as
representative of galactic dust.  We further assume that IG dust differs from
MRN dust due to the removal -- by the selective ejection of large grains and/or
the selective destruction of small grains by sputtering -- of the small-size
end of the grain-size distribution.
In this paper we use absorption and
scattering curves as calculated in Laor \& Draine (1993), and the
truncated MRN distribution with $a_{\rm min}=0.1\mic$; The minimal
grain size (which corresponds to $f\approx 0.4$) is chosen to give
dust which is grey enough not to over-redden the supernovae (A99).
Note that our results should not be sensitive to our adoption of the
Milky Way dust cross section - e.g. LMC dust differs from Milky Way
dust mainly at short wavelengths, where the opacity has anyway been
modified by removing the small grains.
Silicate and graphite grains
are assumed to have equal mass densities, and the total dust density
is chosen, according to the cosmology, as that sufficient to reconcile
the chosen cosmology with the supernova results.  For example,
$\Omega=0.2$ requires $\sim 0.15-0.2$ mag of extinction at $z\sim 0.5$
and hence $\Omega_{\rm dust} \sim (5.5-7)\times 10^{-5}$ to be
consistent with the observations.
\label{sec-dustmodel}

\section{Method of Calculation}

In estimating the contribution from IG dust to the FIRB and the CMB, we follow
the methods of Wright (1981) (see also Loeb \& Haiman 1997).  We define the
comoving number density of photons at redshift $z$ and comoving frequency $\nu$
as
\begin{equation}
N_\nu(z)\equiv\frac{4\pi}{hc(1+z)^3} J_{\nu(1+z)}(z),
\label{eq:photons}
\end{equation}
where $J_\nu(z)$ is the usual specific intensity of the background radiation
field in physical (non-comoving) units of ${\rm
erg~cm^{-2}~s^{-1}~Hz^{-1}~sr^{-1}}$.  The evolution of $N_\nu(z)$ with
redshift is then described by the cosmological radiative transfer equation
\begin{equation}
-\frac{dN_\nu(z)}{dz}=\frac{c\,dt}{dz}\left[ j_{\rm
tot,\nu}(z)-\alpha_\nu(z)N_\nu(z)\right]
\label{eq:transfer}
\end{equation}
where $j_{\rm tot,\nu}=j_{\star,\nu}+j_{\rm dust,\nu}$ is the total comoving
emission coefficient, including both the direct light of UV sources
$j_{\star,\nu}$ (see equation \ref{eq:jstars} below), as well as thermal dust
emission $j_{\rm dust,\nu}$ from dust.  As discussed in section \S~2 above, we
assume that dust is composed of graphite and silicate grains, so that the total
dust absorption coefficient, $\alpha_\nu= \alpha_{\rm Si,\nu}+\alpha_{\rm
Gr,\nu}$ is given in units of ${\rm cm^{-1}}$ by
\begin{equation}
\alpha_\nu(z)=\rho_{\rm Si}(z)\kappa_{\rm Si,\nu(1+z)} + \rho_{\rm
Gr}(z)\kappa_{\rm Gr,\nu(1+z)}
\label{eq:alpha}
\end{equation}
for a total mass density of $\rho_{\rm Si}+\rho_{\rm Gr} =(\Omega_{\rm
Si}+\Omega_{\rm Gr}) \rho_{\rm crit,0} (1+z)^3$ in dust, where the $\kappa$'s
are the dust opacities in units of ${\rm cm^2~g^{-1}}$.  Note that by
assumption, $\Omega_{\rm Si}=\Omega_{\rm Gr}$.
 The total
comoving dust emission coefficient is then
\begin{eqnarray}
j_{\rm dust,\nu} & = & j_{\rm Si,\nu}  + j_{\rm Gr,\nu} \\
                 & = &
\frac{8\pi\nu^3}{c^3}  \{
\frac{\alpha_{\rm Si,\nu}(z)}{\exp[h\nu(1+z)/k_{\rm B}T_{\rm Si}]-1}+
\nonumber \\
& & \frac{\alpha_{\rm Gr,\nu}(z)}{\exp[h\nu(1+z)/k_{\rm B}T_{\rm Gr}]-1}
\}.
\label{eq:jdust}
\end{eqnarray}

The silicate and graphite grains are assumed to be separately in thermal
equilibrium with the total background radiation (CMB + UV sources), so that the
grain temperatures $T_{\rm i}$ are obtained by equating the total amount of
absorption and emission:
\begin{equation}
\int_0^\infty d\nu \,\alpha_{\rm i,\nu} N_\nu(z)=
\int_0^\infty d\nu \,j_{\rm i,\nu},
\label{eq:Tdust}
\end{equation}
where i=Si or Gr.

The infrared background spectrum arises from the re-emission of UV
light absorbed by dust.  The two main factors determining the FIRB are
therefore the evolution of the UV emissivity $j_{\star,\nu}$ from
stellar sources (determining the dust temperature), as well as the
evolution of the dust density, $\Omega_{\rm dust}=\Omega_{\rm
  Si}+\Omega_{\rm Gr}$ -- determining the overall amplitude of the
emissivity. For the UV production rate, here we adopt the global
average star--formation rate as determined recently by Madau et al.
(1998). We ignore any additional UV flux from quasars: at the
redshifts of interest here ($z\lsim 5$), the contribution to the UV
background from the known population of quasars is less than 20\%
(M99).  Note, however, that a population of faint, undetected quasars
could contribute substantially to the UV background (Haiman \& Loeb
1998).

For simplicity, we assume further that the average spectrum from the
stellar UV sources is described as that of black-body radiation at a
temperature of $T=9000$K.  This is a reasonably good approximation to
the continuum in the composite spectra found in the population
synthesis models of Bruzual \& Charlot (1996), although the pure
black-body spectrum is somewhat narrower.  The black-body also falls
below the UV background flux limit at the shortest wavelengths
($\lambda\sim0.1\,\mu$m); see also~\S~\ref{sec-othermod}.
Our results below depend primarily on the total amount of stellar
light absorbed by the dust, and are insensitive to the precise shape
of the UV spectrum (see detailed discussion below). With the above
assumptions, the stellar emissivity is given by

\begin{equation}
j_{\star,\nu}(z) = A_{\rm uv} \dot \rho_\star(z)
B_{\nu(1+z)}(T=9000{\rm K}),
\label{eq:jstars}
\end{equation}

where $B_\nu$ is the Planck function, and $\dot \rho_\star(z)$ is the star
formation rate in Madau et al. (1998), converted from their standard $\Omega=1$
cosmology to the open $\Omega=0.2$ model assumed here, using the multiplicative
factor $(dv_o/d_{l,o}^2)/(dv_\Lambda/d_{l,\Lambda}^2)$, where $d_v$ and $d_l$
are the volume elements and luminosity distances in the two cosmologies
(Eisenstein 1997).

The overall normalization constant $A_{\rm uv}$ is somewhat uncertain. M99
argues that the total UV produced by stars is a factor $\sim 2$ smaller
than the recent measurement of the extragalactic UV background, $\nu J_\nu=20~{\rm
nW~m^{-2}~sr^{-1}}$ at $0.55\,\mu$m by Bernstein (1997).  It is not clear what
the reason for this discrepancy is; the shortage of UV could be provided by
additional faint galaxies or quasars (however, see Vogeley 1997).  Here we take
the simplest approach, and adjust the normalization constant $A_{\rm uv}$ so
that the measured value of $20~{\rm nW~m^{-2}~sr^{-1}}$ for the UV background
at 0.55$\,\mu$m is exactly reproduced (taking into account dust absorption).
Given the shortage of UV provided by the known stellar population, we will also
examine below a model normalized to $10~{\rm nW~m^{-2}~sr^{-1}}$, the lower end
of the $1\sigma$ range quoted by Bernstein (1997).  This also lies above the
lower limits on the UV background derived from galaxy counts in the Hubble Deep
Field (Pozzetti et al. 1998; see figure~\ref{fig-spectrum}). Note that we have
assumed a specific spectral shape and redshift-evolution of the UV
emissivity. We again emphasize that our results are insensitive to the detailed
shape of the UV flux around 1$\,\mu$m, and depend only on the total amount of UV
light absorbed; this is demonstrated by a model we examine below with a more
realistic spectrum obtained from population synthesis models (Bruzual \&
Charlot 1996) and models with SF histories different from M99.  As mentioned
above, the difference between the background inferred from the observed
star-formation rate (i.e.  by direct integration of $\dot\rho_\star$ and an
empirical UV flux per stellar mass ratio), and the measured value of $20~{\rm
nW~m^{-2}~sr^{-1}}$ could be provided by undetected quasars, or other yet
undiscovered sources. As long as the redshift evolution of these sources does
not depart significantly from the range of our assumed $\dot\rho_\star(z)$'s,
our results below would remain valid.

Next, we assume that the same stellar population that produces the UV
background is the source of IG dust.  Under this simple assumption, the mass
density of dust increases in proportion to the mass density in stars:
\begin{equation}
\Omega_{\rm dust}(z)= A_{\rm dust} \int_{t(z=10)}^{t(z)} dt \dot \rho_\star(z)
\label{eq:omdust},
\end{equation}
and we assume further that $\Omega_{\rm Si}=\Omega_{\rm Gr}=
\Omega_{\rm dust}/2$ (as discussed in \S~5 the exact value of this
ratio is not crucial). The normalization constant $A_{\rm dust}$ is
chosen by requiring the total optical depth due to dust absorption +
scattering between $z=0$ and $z=0.5$ to be $\sim 0.15$.  As we argued
in section \S~\ref{sec-dustmodel} above, this is the amount needed to
account for the SN results (A99).  This normalization corresponds to
$\Omega_{\rm dust}(z=0) \sim 5\times10^{-5}$.  Adding to the arguments
given in \S~\ref{sec-dustmodel}, we note that this dust density fits
in comfortably with the amount of dust expected to be produced in
normal stars.  For instance, assuming that each type II SN
($M_\star\geq 8{\rm M_\odot}$) yields $0.3{\rm M_\odot}$ of dust,
adopting a Scalo IMF, and normalizing $\Omega_\star(z=0)=0.004$; we
obtain $\Omega_{\rm dust}(z=0)=5\times10^{-5}$ (cf. Loeb \& Haiman
1997).  This estimate does not account for dust trapped in galaxies or
destroyed, but is also a lower limit since it only includes the dust
formed in supernovae, and not that formed in (for example) dense
clouds or carbon stars.

We note that although stellar dust production cannot occur instantaneously
with star-formation, the relevant time-scales are likely to be much shorter
than the Hubble time at redshifts of interest.  Type II SNe produce dust in a
few million years; while the lifetime of carbon stars (believed to be
dominating the dust budget in the Milky Way, see Gehrz 1989) is still only
$\sim$1 Gyr.  The ejection timescale for dust leaving spiral galaxies is
probably between 100 and 1000 Myr (A99). We have found that incorporating a
delay of $\sim$1 Gyr between star formation and dust production into equation
\ref{eq:omdust} does not change our results below by more than a few percent.

Equations \ref{eq:photons}-\ref{eq:omdust}, together with the initial
condition that $N_\nu$ at some high redshift is equal to the pure CMB with
$T_{\rm cmb}= 2.728(1+z)$K, determine the redshift evolution of the full
background spectrum.
\label{sec-calcmethod}

\section{Existing FIRAS limits}

Any existing IG dust component will process some UV/optical energy into the
FIR/microwave.  As a result, some (perhaps vanishingly small) fraction of the
observed CMB is due to emission from IG dust.  This contribution has been
studied a number of times previously.  Some authors have investigated the
possibility that the CMB is entirely dust-processed emission from astrophysical
objects; see, e.g., Layzer \& Hively (1973), Wright (1982), Wollman (1992) and
Aguirre (1999c) on `cold big-bang' models, or Wickramasinghe et al.
(1975), and Hoyle \& Wickramasinghe (1988) on steady-state models.  In cold
big-bang models, thermalization takes place at very high $z$ after energy
emission by Population III objects.  Both cosmologies require grains of a type
that can maintain a temperature very close to that of the CMB.

If the grains have a temperature only slightly different from that of
the CMB, they will be too cool to contribute to the FIRB; however,
spectral distortions of an initially blackbody CMB will occur (e.g.
Rowan-Robinson, Negroponte \& Silk 1979; Hawkins \& Wright 1988; Bond
et al. 1991).  More recently, this effect has been used by Loeb \&
Haiman (1997) and by Ferrara et al. (1999) to limit the density of IG
dust.  We shall perform a similar analysis using the method and
assumptions outlined in \S~\ref{sec-calcmethod}, and with the dust
model described in \S~\ref{sec-dustmodel}.

Models which distort the CMB at long ($\gg 1$mm) wavelengths can be
easily constrained since there is a clear model for the CMB (perfect
blackbody).  But in the absence of an equally unique and well-defined
model for the FIRB, it is much less clear how to limit FIR emission
that does not actually exceed the observed FIR background.  The
procedure adopted here is as follows: following the methodology of
Fixsen et al. (1996), we fit the results from $500-5000\mic$ by a
blackbody $B_\nu(T_{\rm cmb})$ plus a `uniform dust component'
parameterized by an emissivity and temperature in a $\nu^2B_\nu(T)$
spectrum.  These fitted components are shown in dotted lines in
Figure~\ref{fig-spectrum}. The spectrum with the CMB {\em only}
subtracted is shown as a solid line; this makes clear the contribution
of IG dust emission to the FIRB.  The residuals with both components
subtracted are shown in the second panel, along with the FIRAS
frequency range (vertical dotted lines) and the limit on r.m.s.
deviations from blackbody (horizontal dotted line).

An interesting note about this general procedure is the following. Our
calculations show that the fitted CMB temperature may be close to, but
not exactly the same as the actual initial CMB, and this small
discrepancy affects the measured slope of the FIR emission at $\lambda
\ga 850\mic$, since the FIRB is $\la 1/1000$ of the CMB flux there. For
example, fitting the CMB plus $\nu^2B_\nu$ gives a fitted CMB
temperature of $2.72800$K, exactly as it should be.  But if we fit a
CMB plus $\nu^{0.64}B_\nu$ (the shape of the FIRAS fit), we find
$T_{\rm cmb}=2.72795$K.  The CMB is then slightly under-subtracted, and
the residual FIRB closely follows a $I_\nu \propto \nu^{3.64}$ shape
at long-$\lambda$.  This suggests that accurately determining the
slope of the FIRB long-wavelength tail requires knowledge of $T_{\rm
  cmb}$ to much higher accuracy than available from COBE data.  For
this reason, we find it is not useful to compare the FIRB slope
predicted by models to the FIRAS fit, for $\lambda \ga 850\mic$.

Table~\ref{tab-models} gives the r.m.s.  deviation from blackbody for
the computed spectra, with the mean taken over the $500\mic \le
\lambda \la 5000\mic$ wavelength range analyzed by Fixsen et al.
(1996) and divided by the peak of $B_\nu(T_{\rm cmb})$.  The limit
given by Fixsen et al. 1996 on this number is $5\times 10^{-5}$.  The
result of this analysis is that the dust emission would not cause
detectable distortions to the CMB for the dust model described,
despite the fact that the amount of emission appears to violate the
quoted limits on the $y$ parameter.\footnote{In light of the
  uncertainty in the FIRB, it seems that the quoted limits on r.m.s.
  distortions and the $y$-parameter from the COBE group must be used
  with caution. Rather large spectral distortions can be `hidden' in
  the microwave tail of the FIRB which has a poorly defined spectral
  shape. Assuming conservatively that the FIRB as detected by FIRAS
  exists in the 150-600 GHz range (and zero outside), it corresponds
  to a $y$-parameter in the full 60-600 GHz interval of
  $2\times10^{-4}$, over an order of magnitude above the quoted upper
  limit.}  The residuals are, of course, even smaller if the `uniform
dust component' is fit using a free index in the power law.

On the other hand, modification of the long-wavelength opacity of the dust can
be important, since dust types with higher FIR opacity have lower equilibrium
temperatures which may lead to excessive emission into the FIRB/CMB.  Since the
dust emission tends to be well described by the fitted $\nu^2B_\nu$ isotropic
dust component, this emission does not tend to leave excessive distortion in
the CMB residuals.  Rather, it leads to a fit of the isotropic dust component
incompatible with that found by FIRAS. Accordingly, we discuss these models in
the next section, which treats the dust contribution to the FIRB and the
resulting limits on the dust models.
\label{sec-cmbdist}

\section{IG Dust Contribution to the FIRB}

In this section, we analyze the contribution by IG dust emission to the FIRB in
a variety of models.  Section~\S~\ref{sec-fidmod} presents and discusses our
fiducial model.  Section \S~\ref{sec-othermod} lists possible variations in the
parameters, and discusses the effects of these variations, and section
\S~\ref{sec-resolve} discusses how resolution of the FIRB into discrete sources
by future experiments would improve the constraints.

\subsection{Fiducial Model}

In our fiducial model, the cosmology is open, with $\Omega=0.2$ and
$h_{50}=1$.  The star formation rate is taken from M99 and shown in
A99, Figure 1. The dust has equal parts by mass in silicate and
graphite grains with $a_{\rm min}=0.1\mic$, and density proportional
to the integrated SFR, normalized to $\Omega_{\rm dust}(z=0) =
5.4\times 10^{-5}$, which gives 0.15 mag of extinction to $z=0.5$ at
(observed) $0.66\mic$.  The galaxy spectrum is a 9000 K blackbody,
normalized to give $20\iun$ after dust processing at $0.55\mic$.

The predictions of the fiducial model appear in
Figure~\ref{fig-spectrum} and Table~\ref{tab-models}.  The figure
shows the complete final spectrum with the CMB (fit using the method
of \S~\ref{sec-cmbdist}) subtracted.  For comparison we also include
the UVB lower limits (Pozzetti et al. 1998) and detections (Bernstein
1997), the DIRBE FIRB upper limits and detections (Hauser et al.
1998), and the FIRAS detections (Fixsen et al. 1998) with $\pm
1\sigma$ uncertainties.  It is apparent that the dust emission
contributes significantly to the long-wavelength flux -- in fact it
can account for all of the $850\mic$ flux given by the FIRAS
measurements -- but fails to reproduce the shape or amplitude of the
entire FIRB.  This is specified quantitatively in
Table~\ref{tab-models}, where the fifth and sixth columns list the
numerical fraction of the FIRAS FIRB measurements that the dust
emission would contribute at $200$ and $850\mic$.  These numbers
should be multiplied by 0.76 to yield the fraction of the 1$\sigma$
upper limit on the FIRB, or 1.25 to account for the minimum $20\%$
resolved fraction.  Column seven lists the $y$-parameter represented
by the FIR emission, defined by $y\equiv 0.25(u_{\rm tot}/u_{\rm
  cmb}-1)$, where $u_{\rm tot}$ and $u_{\rm cmb}$ are the total energy
density and the energy density of the CMB alone, both evaluated in the
60-600 GHz frequency range.  This gives a measure of the total dust
energy output.  The last column gives the r.m.s.  residual after both
CMB and the $\nu^2B_\nu$ fit to the FIR emission are subtracted, as
discussed in \S~\ref{sec-cmbdist}.

The next section analyzes the possible variations upon these predictions
resulting from changes in the model.  Some of the variations can be ruled out
because they predict a FIRB contribution above the FIRAS limit; these models
could be saved only if the FIRAS team has over-subtracted the foreground
contribution.  Stronger limits can be obtained on the various models by
considering their contribution to the {\it unresolved} fraction of the FIRB,
after the contributions by known discrete sources are subtracted.  These
constraints will be discussed in section~\ref{sec-resolve} below.

\label{sec-fidmod}

\vspace{.5in}
\begin{deluxetable}{lcccccccc}
\tablecaption{Intergalactic dust emission models}
\tablewidth{0pt}
\tablehead{
\colhead{Model}   &
\colhead{Variation}        &
\colhead{$T_{\rm Gr}$} &
\colhead{$T_{\rm Si}$} &
\colhead{$f_{200}$}    &
\colhead{$f_{850}$}    &
\colhead{$y_{\rm FIRB}/10^{-4}$} &
\colhead{rms$/10^{-6}$}}
\startdata
Fid. & - & 9.9 & 7.3 & 0.14 & 1.02 & 2.0 & 3.7 \nl
H80 & $H_0=80{\rm\,km\,s^{-1}\,mpc^{-1}}$ & 9.9 & 7.3 & 0.15 & 1.10 & 2.1 & 4.0 \nl
FSFR & SFR flat for $z > 3$ & 9.9 & 7.3 & 0.16 & 1.45 & 2.7 & 7.4 \nl
BC & Galaxy spectra from Bruzual-Charlot & 9.3 & 7.8 & 0.08 & 0.89 & 1.7 & 2.9
\nl
BC2 & Bruzual-Charlot + extreme reddening & 10.5 & 7.9 & 0.24 & 1.26 & 2.5 & 4.6 \nl
MG & $a=0.1\mic$ grains, $\Omega_{\rm d}=4.5\times 10^{-5}$
& 10.2 & 7.1 & 0.17 & 0.96 & 1.9 & 3.4 \nl
LG & $a=0.25\mic$ grains, $\Omega_{\rm d}=7.6\times 10^{-5}$
& 9.2 & 7.4 & 0.08 & 0.85 & 1.6 & 3.1 \nl
LUV & UVB of $10\iun$ at $0.55\mic$ & 8.8 & 6.5 & 0.05 & 0.72 & 1.3 & 2.5 \nl
SCDM & $\Omega=1$, $\Omega_{\rm d}(z=0) = 1.25\times 10^{-4}$ & 9.9 & 7.3 & 0.30 & 1.94 & 3.8 & 6.7 \nl
SCDMb & SCDM, but UVB of $10\iun$ & 8.8 & 6.5 & 0.10 & 1.33 & 2.5 & 4.4 \nl
G05 & $\kappa$ flat to $\lambda_0=0.5\mic$ & 17.3 & - & 0.86 & 0.45 & 1.1 & 1.1 \nl
G2 & $\kappa$ flat to $\lambda_0=2\mic$ & 12.8 & - & 0.88 & 1.56 & 3.5 & 4.6 \nl
G5 & $\kappa$ flat to $\lambda_0=5\mic$ & 9.7 & - & 0.53 & 3.97 & 7.8 & 12 \nl
G10 & $\kappa$ flat to $\lambda_0=10\mic$ & 7.7 & - & 0.24 & 7.71 & 13.4 & 19 \nl
L1.0 & $\kappa(\lambda \ge 100\mic) \propto \lambda^{-1}$ &7.9 & 5.5 & 0.038 & 2.59 & 4.4 & 18 \nl
L1.5 & $\kappa(\lambda \ge 100\mic) \propto \lambda^{-1.5}$ & 9.0 & 6.4 & 0.083 & 1.71 & 3.1 & 9.5 \nl
L2.5 & $\kappa(\lambda \ge 100\mic) \propto \lambda^{-2.5}$ & 10.6 & 8.0 & 0.19 & 0.65 & 1.3 & 1.7 \nl 
GRA & graphite, $a_{\rm min}=0.05\mic$ & 9.8 & - & 0.21 & 1.29 & 2.6 & 4.4 \nl
S2 & 1:2 graphite: silicate by mass & 9.9 & 7.3 & 0.12 & 1.00 & 1.9 & 3.7 \nl
\enddata
\label{tab-models}
\end{deluxetable}

\subsection{Variations on the model}

Several variations were performed to test the robustness our model
assumptions.  First, raising $H_0$ from $50$ to
$80{\rm\,km\,s^{-1}\,mpc^{-1}}$ (model H80) made very little
difference.  Second, an SFR which is flat for $z > 3$, but unchanged
for $z<3$ was employed (model FSFR).  This choice is motivated by the
``effective'' global average SFR (M99) required to reionize the
universe by $z=5$.  This increases the dust contribution to the final
spectrum by about 35\%, and the $850\mic$ contribution by 45\%. The
change occurs because relatively more of the UV energy is emitted at
high $z$, where the dust is more dense, and emits into longer
(rest-frame) wavelengths.

Next, we tested the adequacy of our assumption of the
single-temperature blackbody galaxy spectrum, by using a more
realistic spectral template.  We utilize the Bruzual-Charlot (1996)
starburst population synthesis model with a Scalo IMF to compute the
shape of spectrum as a function of the age of the population. This
time-dependent template is convolved with the SFR from M99 to find the
redshift-evolution of the frequency dependent UV emissivity.  The full
spectrum under these assumptions is shown by the curve labeled ``BC''
in Figure~\ref{fig-spectra1}.  The peak in the UVB from the direct
galactic emission is broader than in our fiducial model, and extends
to higher energies.  As mentioned above, this procedure results in a
UVB at $z=0$ that falls short of the measured value by a factor of
$\sim 2$.  Overall, the contribution to the FIRB is not effected
significantly: it is reduced only by $\sim 10\%$ relative to our
fiducial model at $850\,\mu$m (cf. Table~\ref{tab-models}).

It must be noted that the above procedure is not strictly correct, as
we have not corrected the galactic templates for reddening by dust
internal to the galaxy.  This reddening could be important in the
present context, since it makes the UV peak steeper in the
$0.1-1\,\mu$m range, and allow a broader UVB in this wavelength range
that reaches $\sim 20\iun$ at $0.55\,\mu$m, while extending to
energies higher than assumed in our fiducial model.  Here we do not
attempt to derive an accurate reddening correction (see instead Madau
et al. 1998).  Rather, our aim is to assess the maximum UV flux that
could, in principle, be present in the $0.1-1\,\mu$m range, that is
consistent with both the Voyager upper limit (Murthy, Henry \& Holberg
1998) and the $0.55\,\mu$m detection.

To compute this maximal UV flux, we reddened the galactic template
spectrum, assuming the cross section of Milky Way dust (DL), with a
total (unrealistically high) maximum optical depth of $\sim4$.  We
then re-scaled the UV efficiency by a factor of $\sim$10 to retain an
intensity of $\sim 20\iun$ at $0.55\,\mu$m.  The resulting spectrum is
shown by the curve labeled ``BC2'' in Figure~\ref{fig-spectra1}.  This
spectrum fits in with both the Voyager and Bernstein (1997) data
points, and reasonably represents the maximally allowed UV emission.
It is interesting to note that a simultaneous fit to these two
data-points requires a very steep spectrum ($20\iun$ at 0.55$\,\mu$m
to $0.6\iun$ at 0.1$\,\mu$m, implying a slope of $I\sim \lambda^2$) ,
which can only be derived from the Bruzual-Charlot template by
postulating an exceedingly high reddening.  One resolution of this
paradox might be that the true value of the UVB is closer to $10\iun$
(as argued by Vogeley 1997 in a different context).  We find that with
this maximal emission, the FIRB contribution increases somewhat, but
the change is once again not significant, only $\sim 40\%$ relative to
the galactic model without reddening, over-predicting the 850$\,\mu$m
FIRB by $\sim 26\%$ (cf. Table~\ref{tab-models}), but still well
within the $+1\sigma$ uncertainty.

Third, we have tested our assumption that we may treat the grain-size
distribution as a single component, by considering grains of a single
grain size $a$, for $a=0.1\mic$ (model MG) and $a=0.25\mic$ (model
LG).  These extremes change the $z=0$ temperature by $0.1-0.7\K$ and
the $850\mic$ flux by $\la 4-15\%$, demonstrating that the
approximation is not very important.  The test also shows that the
results are not sensitive to the grain size distribution, as long as
all of the grains are relatively large ($a \ga 0.1\mic$).  The exact
graphite:silicate ratio is also fairly unimportant; models GRA and S2
have graphite:silicate ratios of 1:0 and 1:2 respectively.  The former
corresponds to a scenario in which silicates are preferentially
destroyed (see A99); the latter is an extreme of reasonable
graphite:silicate ratios for two-component model, as shown in the
recent computations of Weingartner \& Draine 1999.  Neither is
drastically different from the fiducial model, and the S2 model is
almost identical.

Finally, we check our assumption that the dust temperature follows from
equilibrium with a homogeneous background radiation bath, because dust near
concentrations of radiation could have somewhat higher temperature.  To
determine whether this might be important, we have computed the critical
distance $r_{\rm crit}(z)$ (from center of a model galaxy) at which a dust
grain would feel equal contributions from the cosmic UVB intensity $\nu J_\nu(z)$
and from the nearby galaxy, i.e.
\begin{equation}
4\pi (1+z)^3 \nu J_\nu = {L_{\rm gal}\over 4\pi r_{\rm crit}^2},
\end{equation}
where the galaxy luminosity $L_{\rm gal} \sim
10^{46}{\rm\,erg\,s^{-1}}$ (corresponding to a $10^{12}\msun$
starburst galaxy, with the spectral model of Bruzual \& Charlot 1996)
is constant and $\nu J_\nu(z)$ is proportional to the integrated
comoving SFR and has the value $20\iun$ at $z=0$.  The result is that
$r_{\rm crit} \la 60\,$kpc for all $z$ and $r_{\rm crit} \la 30\,$kpc
for $z > 1$.  In A99, it was argued that IG dust must be $\ga 70\,$kpc
from its progenitor galaxy to be uniform enough to be in accord with
limits on the dispersion in supernova brightnesses.  Thus even near
extremely bright galaxies, dust of the A99 model should have
temperature dominated by the isotropic background rather than nearby
galaxies.

As discussed in~\S\ref{sec-calcmethod}, there are still sizeable
uncertainties in the normalization of the UV spectrum.  The 1$\sigma$
lower limits from Bernstein (1997) roughly coincide with the lower
limits given by integrated counts (M99) at $\sim 10\iun$ at
$0.55\mic$.  We have employed this normalization (model LUV), and find
that this lowers the $850\mic$ flux by about 30\% (the decrease in
energy is somewhat offset by the lower temperature of the dust).

Although most models considered predict an $850\mic$ flux comparable
to that detected by FIRAS, some models exceed the FIRAS upper limit.
To reconcile the supernova measurements with a closed,
matter-dominated cosmology -- such as the standard cold dark matter
model (SCDM) -- an extinction of $A_V(z\sim0.5)\sim 0.35{\rm\,mag}$ is
required (A99), in turn requiring $\Omega_d(z=0) \approx 1.25\times
10^{-4}$ for $h_{50}=1.$ In the present model this predicts an
$850\mic$ flux which exceeds the $1-\sigma$ bound of the FIRAS
detection (see the SCDM model in Table~\ref{tab-models}.)  With the
lower UV normalization, however, the emission is 
marginally compatible with the errors.
(model SCDMb in Table~\ref{tab-models}).  If the higher UV
normalization holds, these results make it very unlikely that IG
dust could reconcile SCDM with the supernova observations, even if the
large necessary quantity of dust existed (cf. A99).

The present calculations can also constrain the properties of
intergalactic dust if it is assumed to be responsible for the
supernova dimming.  Very grey dust, for instance, absorbs more
UV/optical flux for a given V-band extinction (and can hence
overproduce the FIRB).  Both very grey dust and `standard' dust with
high IR emissivity can also have lower equilibrium temperature, which
overproduces the FIRB at long wavelengths.  To limit these sorts of
dust, we have considered two illustrative types.  The first is dust
that is grey out to some wavelength $\lambda_0$, then falls as
$\lambda^{-2}$:
\begin{equation}
\kappa_\lambda \propto {1 \over [1+(\lambda/\lambda_0)^2]}.
\label{eq-needlekappa}
\end{equation}

This is conservative in the sense of ensuring the maximum reasonable
temperature for a given $\lambda_0$.  Models (G05-G10) are of this
type, and show that for $\lambda_0 \ga 3\mic$ the FIRB is
overproduced.  It is interesting to note, however, that $\lambda_0
\approx 0.5$ models actually produce a FIRB fitting the observations
rather well over the entire frequency range.  Conducting needles have
an absorption spectrum which can be roughly modeled by
eq.~\ref{eq-needlekappa} (Wright 1982; Wickramasinghe \& Wallis 1996;
Aguirre 1999a), and using Wright's 1982 RC model with a resistivity of
$\sim 10^{-15}{\rm\,s}$, we can rule out needle models with
length-to-diameter ratio $L/d \ga 6$.\footnote{Needles with high $L/d$
  and high conductivity, such as the iron whiskers of Hoyle \&
  Wickramasinghe (1988), can maintain a temperature very close to that
  of the CMB and are not constrained by the present calculations; see
  also Aguirre (1999c).}

\begin{figure}
\plotone{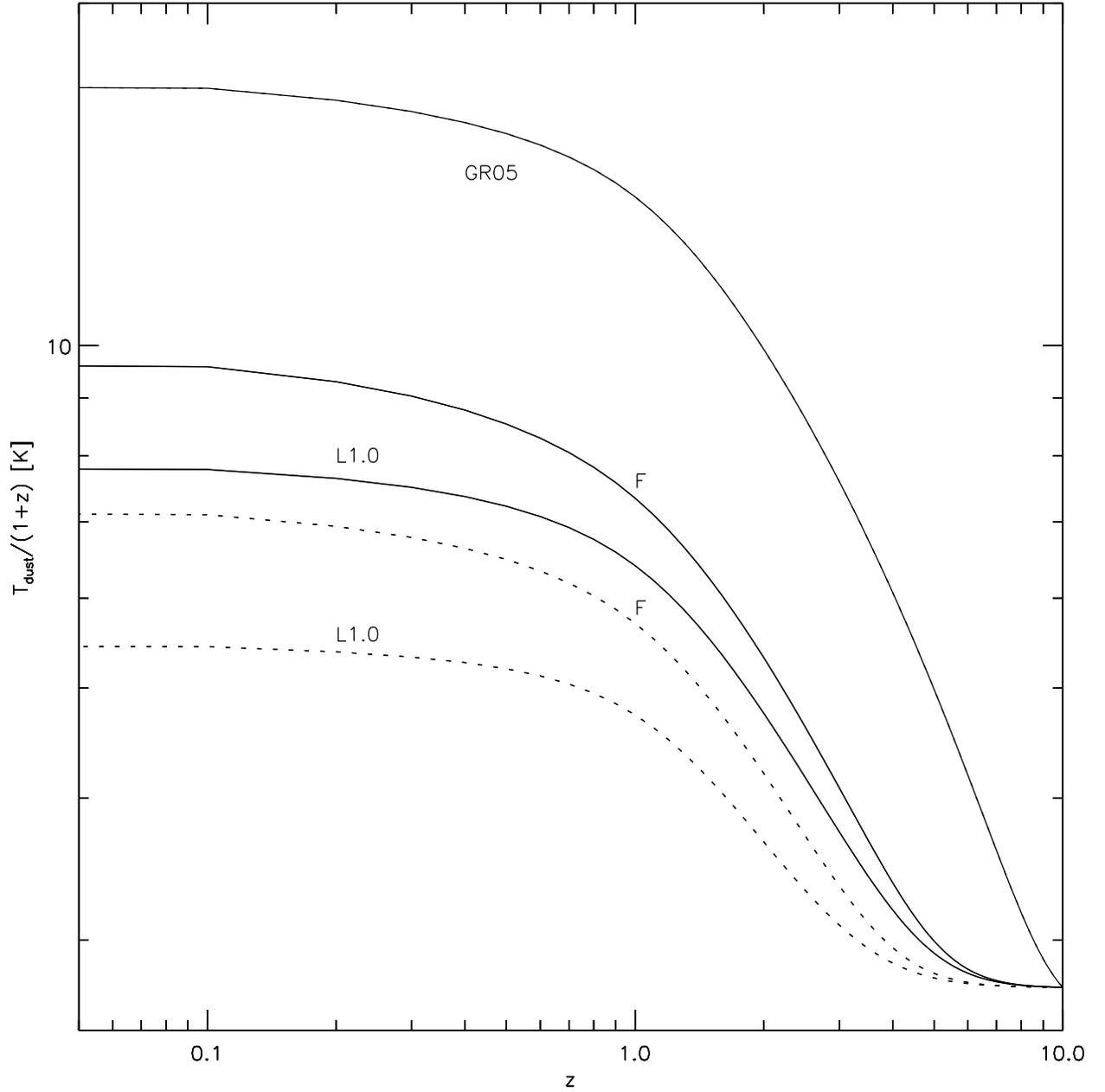}
\caption{Evolution of dust temperature for silicate (dotted) and graphite (solid)
  grains in our fiducial model (F), a dust model with long-wavelength
  opacity $\propto \lambda^{-1}$ (L1.0) and a model with opacity that
  is constant for $\lambda < 0.5\mic$ and falls as
  $\lambda^{-2}$ for longer $\lambda$ (GR05).  The dust temperatures in
  all other models fall in-between the two extreme models shown here.}
\label{fig-tdust}
\end{figure}

\begin{figure}
\plotone{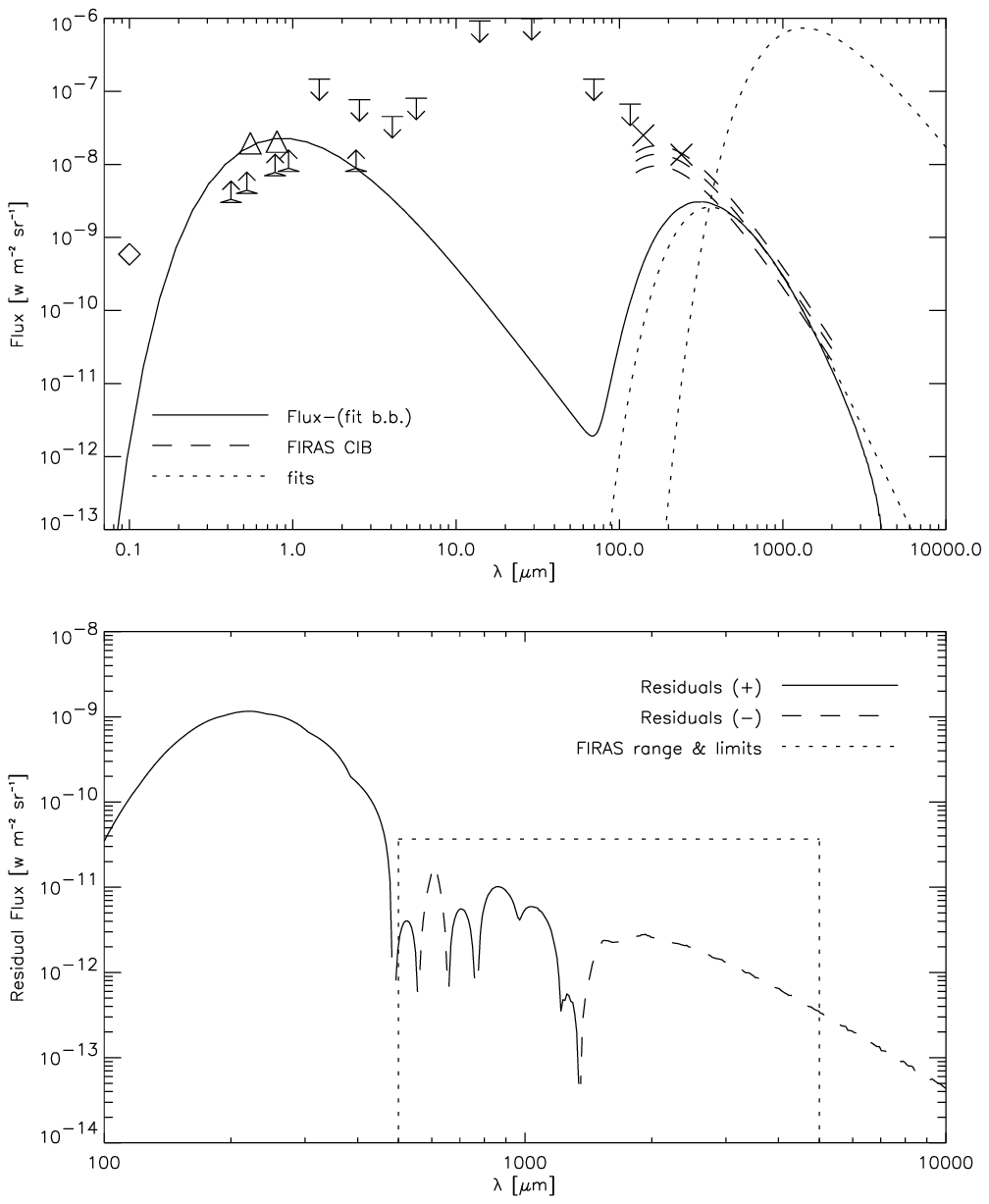}
\caption{{\bf Top:} Full $z=0$ output spectrum (solid line) for our
 fiducial model, with the best-fit CMB (higher dotted line) subtracted.
  Also shown are the fitted $\nu^2B_\nu$ component (lower dotted line),
  the FIRAS detected FIRB $\pm1\sigma$ bounds (dashed), DIRBE FIRB
  detections (crosses) and upper limits (down arrows), the UVB detections
  (triangles) and HDF-derived lower limits (up arrows), and the
  Voyager FUV upper limit (diamond).  {\bf Bottom:} Spectrum with both
  CMB and $\nu^2B_\nu$ component subtracted (solid and dashed lines).
  Vertical dotted lines represent the wavelength range over which the
  FIRAS distortion limit analysis was performed, the horizontal
  dotted line indicated the r.m.s. of the residuals in the FIRAS
  data. The features in the 500-5000$\,\mu$m range indicate the level
  of numerical noise in our calculations.}
\label{fig-spectrum}
\end{figure}

\begin{figure}
\plotone{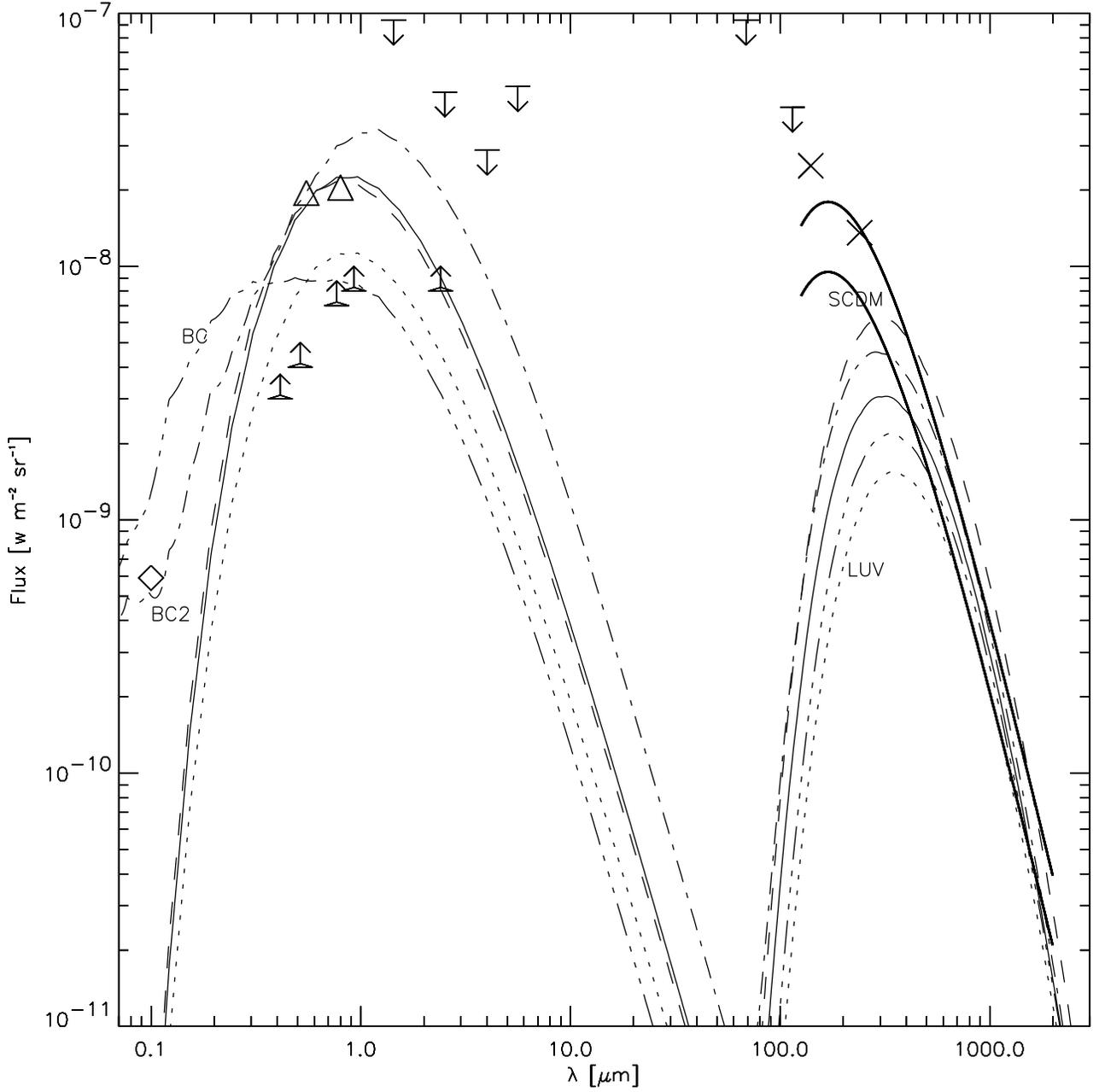}
\caption{Output spectra for various models: A model
  using half the UVB normalization (LUV; dotted), an $\Omega=1$ model
  (SCDM; dashed), and models with the Bruzual-Charlot spectral
  templates (BC; triple-dot-dashed), and reddened B-C templates (BC2;
  dot-dashed).  The fiducial model is included (solid), as are UVB
  and FIRB limits as in Figure~\ref{fig-spectrum}, with FIRAS FIRB
  detection limits in dark, solid.  See text and
  Table~\ref{tab-models} for details on models.}
\label{fig-spectra1}
\end{figure}

The second dust type is `standard' DL/MRN dust, but with
long-wavelength ($\lambda \ga 100\mic$) emissivity given by
$\kappa_\lambda \propto \lambda^{-\alpha}$ with $\alpha \neq 2.$
Fluffy/fractal grains can give $\alpha \sim 1$ (e.g. Wright 1987;
Stognienko et al. 1995), and $\alpha$ generally depends upon the
optical properties of the grain material.  Dust with $\alpha < 2$
generally has lower temperature, and the calculations (see models
L1.0, L1.5 in Table~\ref{tab-models}) show that such dust will
overproduce the $850\mic$ FIRB (at the $1\sigma$ level) for $\alpha
\la 1.5$.  On the other hand, if dust had $\alpha > 2$, model L2.5
shows that the higher resulting temperature would lead to less flux at
$850\mic$.  The comoving dust temperature $T_{\rm dust}/(1+z)$ for the
fiducial model and models GR05 and L1.0 is shown in
Figure~\ref{fig-tdust}.  All computed models fall within the range
spanned by the GR05 and L1.0 models.

In summary, using only the FIRAS limits on the FIRB, we can with some
confidence rule out SCDM models, `very grey' dust models with
$\lambda_0 \ga 3\mic$, and dust models with lower law index $\la 1.5$
in their long-wavelength emissivity.  The next section discusses
stronger constraints which derive from the fraction of the measured
FIRB which is known to be due to discrete sources.
\label{sec-othermod}

\subsection{Resolving the FIRB: Tests and future observations}

When constraining models using their predicted IG dust contribution to
the FIRB, we have so far considered the full value of the FIRB.
However, as mentioned above, we already know that the entire FIRB
cannot be comprised of IG dust emission alone. On the one hand, the IG
dust models generally fail to produce the short wavelength fluxes; on
the other hand, observations have already resolved a non-negligible
fraction $(\gsim 25\%)$ of the FIRB into discrete sources.  If
$20-30\%$ of the FIRB is accounted for by resolved sources, the
numbers in column six of Table~\ref{tab-models} should be multiplied
by 1.25-1.5, as only $70-80\%$ of the measured FIRB can be IG dust
emission.  To use instead the $1\sigma$ upper limit on the detected
FIRB, but taking into account the resolved fraction, column six should
be multiplied by $0.95-1.14$.  The calculations therefore indicate
that resolving the FIRB further could provide significantly stronger
constraints on the models.  It must also be noted, however, that there
is a $\sim 10-20\%$ uncertainty in the SCUBA flux calibration, and
that the quoted errors on the measured FIRB are really rough estimates
of systematic uncertainty reflecting only differences between the fits
resulting from different foreground subtraction method.  For
comparison, Guiderdoni et al. (1997) give an allowed FIRB intensity
range at $\sim 850\mic$ of $\sim 0.2-1.5\times10^{-9}\iun$ at
$1-\sigma$. These uncertainties correspond to roughly four to five
times those quoted by Fixsen et al.

At the present time, a handful ($\sim 30$) of objects with fluxes $S$
above $S\geq 2\,$mJy at $850\mic$ have been detected by SCUBA.  These
data have been used to derive the empirical number counts $n(S)$ of
sources, as a function of apparent flux.  At the limiting flux of
$S=2\,$mJy, the total contribution from these objects is a fraction
$20-30\%$ of the FIRB value (Barger et al. 1999a).  As emphasized by
these authors, the faint-end counts are very steep ($n(S)\propto
S^{-3.2}$).  Extrapolating at the faint end of the number counts to
$\sim 0.5\,$mJy would then account for the entire measured background.
These conclusions are further supported by the somewhat less reliable
detections of faint sources down to $\approx 0.75\,$mJy: the gravitationally
lensed sample of Blain et al. (1999a,b) extends to this limit, albeit
with only a four sources fainter than $\sim 2\,$mJy.

In light of these existing measurements, and especially because of the
steepness of the faint-end number counts, tightening the constraints
obtained in this paper requires relatively small further improvements
in the observations.  In particular, with a faint-end slope of -3.2,
the flux limit needs to be lowered only by a factor of four, to $\sim
0.5\,$mJy, to detect the sources that account for the whole of the
FIRB, provided these sources exist.  If the faint sources do not
exist, then these measurements will find the turn-over and flattening
in $n(S)$, and yield the actual fraction of the FIRB not accounted for
by these sources.  If the faint sources are detected, the uncertainty
in their total cumulative flux would be relatively small, since these
sources are expected to be numerous.

A complication we have ignored in this analysis is the following: most
IG dust emission occurs at $1 \la z \la 3$, where a tentative redshift
analysis (Barger et al. 1999b and references therein) indicates that
most of the SCUBA $850\mic$ sources lie.  The source fluxes are
derived using a 30 arcsec aperture, which at $z > 1$ corresponds to a
physical size of $\ga 250h_{65}^{-1}\,$kpc. We have assumed for
simplicity that the IG dust is effectively uniform, but at $z>1$
significant quantities could be within $125\,$kpc of galaxies, so
there is a danger that the SCUBA derived fluxes include some emission
from what we have termed IG dust.  Upcoming sub-mm surveys with better
angular resolution should be able to clarify this ambiguity.

In summary, with a factor of four improvement over the current SCUBA
detection threshold, we would either see explicitly the need for
something else to explain the FIRB, or else a fairly tight constraint
on the IG dust will be possible.  Forthcoming instruments such {\it
  Bolocam} and {\it SIRTF} will reach the required sensitivities at
160$\,\mu$m and 1100$\,\mu$m (see Table 1 in Blain 1999).  The constraints
that can be obtained then might then be limited by the accuracy of the
absolute measurement of the coarse-grained FIRB.  The High Frequency
Instrument (HFI) on the future Planck satellite\footnote{See
  http://astro.estec.esa.nl/SA-general/Projects/Planck.}, covering the
$350-3000\,\mu$m range at several intermediate frequencies, will be able
to greatly improve our existing knowledge of the unresolved component.
Finally, as shown by Haiman \& Knox (1999), angular correlations in the
FIRB are expected to be at the few percent level, and depend strongly
on the nature of its sources.  Future instruments such as {\it
BLAZE} and {\it FIRBAT} will be able to measure these correlations,
providing another diagnostic that distinguishes between IG dust and
discrete sources.

\label{sec-resolve}

\section{conclusions}

In this paper, we studied the far infrared emission from the type and
amount of IG dust necessary to explain the recent Hubble diagrams,
derived from observations of Type Ia SNe at redshifts $0<z<1$, in
cosmological models without a cosmological constant.  In particular,
we computed the contribution from the IG dust emission to the value of
the FIRB recently measured by the COBE satellite.  We investigated a
broad range of models, and focused on the wavelength of $850\,\mu$m,
where the largest fraction of the FIRB is presently resolved into
discrete sources, thus yielding the strongest constraints.  The
current constraints are limited by $\sim 10-30\%$ errors in each of the
following: the dust model, the measured FIRB flux, the SCUBA counts
and the SCUBA calibration.

Our results show that the IG dust emission is consistent with the
spectral distortion of the CMB allowed by present COBE data, but may
contribute nearly all of the {\it unresolved} fraction of the FIRB in
the $300-1000\,\mu$m range.  In a few specific models, this
contribution is sufficiently large to render those models implausible,
including Standard CDM and models with very grey dust, or dust with
unusually high IR emissivity.  In the (perhaps most interesting) case
of an open universe with a low matter density ($\Omega=0.2$), we find
that the contribution is still within the experimental uncertainty.
Assuming that 20\% of the FIRB is accounted for by the discrete SCUBA
sources, the IG dust contributes up to $\sim 90\%$ of the $+1\sigma$
limit on the {\it unresolved} fraction.  Future observations of the
far infrared background by Planck, and its discrete constituents by
{\it SIRTF} and {\it Bolocam} will provide a definitive test of the IG
dust hypothesis.

\acknowledgments

We thank I. Smail, R. Ivison and D. Fixsen for helful communications,
and thank P. Madau for providing a fitting formula for the SFR.  ZH
was supported by the DOE and the NASA grant NAG 5-7092 at Fermilab.
This work was supported in part by the National Science Foundation
grant no. PHY-9507695.

\clearpage
\newpage

\clearpage
\newpage

\clearpage
\newpage

\clearpage
\newpage
\begin{figure}
\plotone{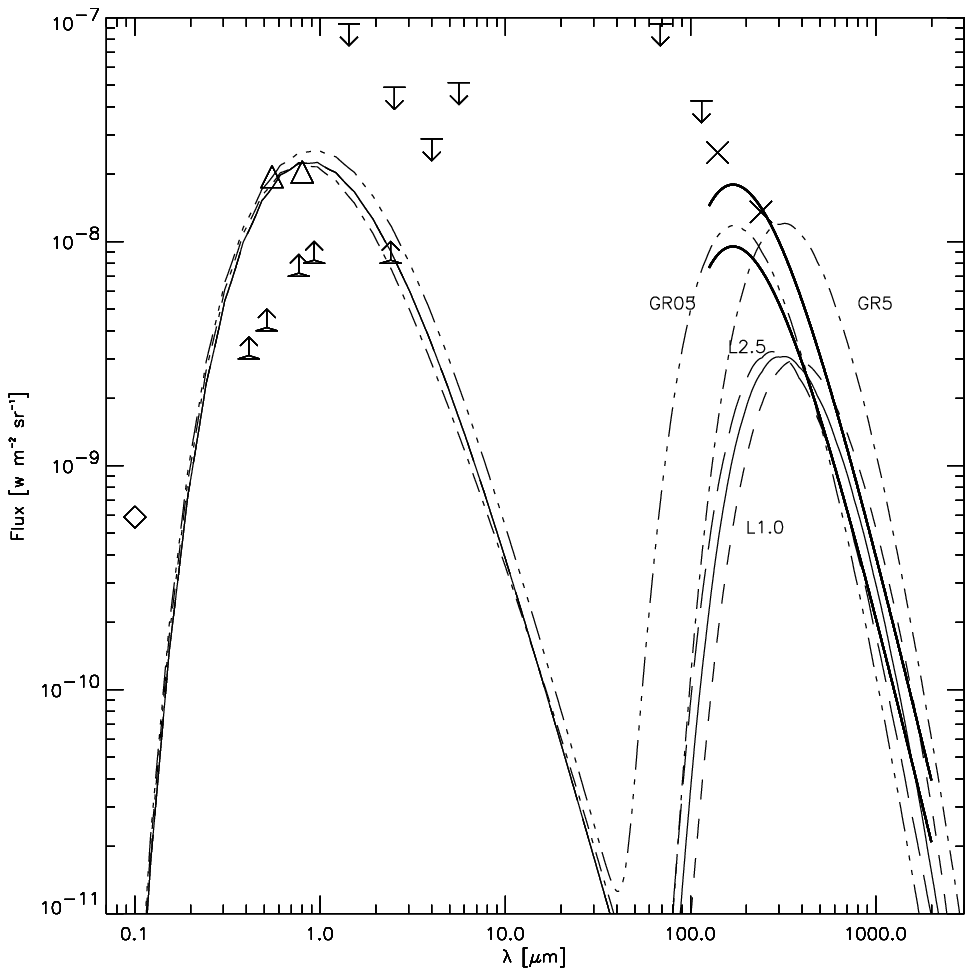}
\caption{Output spectra for different dust models: A model
  with long-wavelength opacity $\propto \lambda^{-\alpha}$ for
  $\alpha=1$ (L1.0; dashed) and $\alpha=2.5$ (L2.5; long dashed), and
  models with dust which is grey out to $\lambda_0$ then falls as
  $\lambda^{-2}$, for $\lambda_0=0.5\mic$ (GR05; triple-dot-dashed)
  and for $\lambda_0=5\mic$ (GR5; dot-dashed).  Again, the fiducial
  model is included (solid), as are UVB and FIRB limits as in
  Figure~\ref{fig-spectrum}, with FIRAS FIRB detection limits in dark,
  solid.  See text and Table~\ref{tab-models} for details on models.}
\label{fig-spectra2}
\end{figure}


\begin{thebibliography}{99}
\bibitem[]{a99} Aguirre, A. 1999a, ApJ, 512, L19
\bibitem[]{a99b} Aguirre, A. 1999b [A99], ApJ, 525, in press; 
astro-ph/9904319
\bibitem[]{a99c} Aguirre, A. 1999c, ApJ, submitted
\bibitem[]{barger99} Barger, A. J., Cowie, L. L., \& Sanders, D. B. 1999a,
 ApJL, in press, astro-ph/9904126
\bibitem[]{barger99b} Barger, A. J., Cowie, L. L., Smail, I., Ivison, R. J.,
 Blain, A. W. \& Kneib, J.-P. 1999b, AJ, in press; astro-ph/9903142
\bibitem[]{bernstein97} Bernstein, R. 1997, PhD thesis, Caltech
\bibitem[]{blain99a} Blain, A. W. 1999, in Proc. of Photometric Redshift Meeting, OCIW, April 1999, preprint astro-ph/9906141
\bibitem[]{blain99b} Blain, A. W., Kneib, J.-P., Ivison, R. J., \& Smail, I. 1999a, ApJ, 512, L87
\bibitem[]{blain99c} Blain, A. W., Ivison, R. J., Kneib, J.-P., \& Smail, I. 1999b, astro-ph/9908024
\bibitem[]{bond91} Bond, J. R., Carr, B. J., \& Hogan, C. J. 1991, ApJ, 367, 420
\bibitem[]{bruzual96} Bruzual, G., \& Charlot, S. 1996, in preparation.  The models are available from the anonymous ftp site gemini.tuc.noao.edu.
\bibitem[]{dl:a} Draine, B. \& Lee, H. 1984, {ApJ}, {285}, 89
\bibitem[]{eisenstein97} Eisenstein, D. J. 1997, ApJ, submitted, preprint astro-ph/9709054
\bibitem[]{1999MNRAS.303..301F} Ferrara, A., Nath, B., Sethi, S. K. \& Shchekinov, Y. 1999, \mnras, 303, 301
\bibitem[]{1996ApJ...473..576F} Fixsen, D. J., Cheng, E. S., Gales, J. M., Mather, J. C., Shafer, R. A. \& Wright, E. L. 1996, \apj, 473, 576
\bibitem[]{1998ApJ...508..123F} Fixsen, D. J., Dwek, E., Mather, J. C., Bennett, C. L. \& Shafer, R. A. 1998, \apj, 508, 123
\bibitem[]{gehrz} Gehrz, R. D. 1989, in ``Interstellar Dust", Proc. of the 135th IAU Symposium, held in Santa Clara, California, 26-30 July, 1988, Eds. Louis J. Allamandola and A. G. G. M. Tielens, Kluwer Academic Publishers, Dordrecht, p.445
\bibitem[]{guiderdoni97} Guiderdoni, B., Bouchet, F. R., Puget, J.-L., Lagache, G., \& Hivon, E. 1997, \nat, 390, 257
\bibitem[]{guiderdoni98} Guiderdoni, B., Hivon, E., Bouchet, F. R., \& Maffei, B. 1998, MNRAS, 295, 877
\bibitem[]{haiman99} Haiman, Z. \& Knox, L. ApJL, submitted, preprint astro-ph/9906399
\bibitem[]{hl98} Haiman, Z. \& Loeb, A. 1998, \apj, 503, 505
\bibitem[]{1998ApJ...508...25H} Hauser, M. G., et al. 1998, \apj, 508, 25
\bibitem[]{wright:b} Hawkins, I. \& Wright, E. 1988, ApJ, 324, 46
\bibitem[]{hoyle:a} Hoyle, F. \& Wickramasinghe, N. 1988, Ap\&SS, 147, 245
\bibitem[]{1998Natur.394..241H} Hughes, D. H., et al. 
1998, \nat, 394, 241
\bibitem[]{draine:b} Laor, A. \& Draine, B. 1993, {ApJ}, {402}, 441
\bibitem[]{layzer:b} Layzer, D. \& Hively, R. 1973, ApJ, { 179}, 361
\bibitem[]{loeb97} Loeb, A. \& Haiman, Z. 1997, 490, 571
\bibitem[]{madau99} Madau, P. 1999, to appear in Physica Scripta, Proceedings of the Nobel Symposium, Particle Physics and the Universe (Enkoping, Sweden, August 20-25, 1998), astro-ph/9902228
\bibitem[]{madau98} Madau, P., Pozzetti, L. \& Dickinson, M. 1998, ApJ, 498, 106
\bibitem[]{mathis:c} Mathis, J., Rumpl, W. \& Nordsieck, K. 1977, ApJ, 217, 425
\bibitem[Murthy Henry \& Holberg 1998]{1998AAS...193.6509M} Murthy, J., 
Henry, R. C. \& Holberg, J. B. 1998, American Astronomical Society Meeting, 
193, 6509
\bibitem[]{pei:a} Pei, Y., Fall, M. \& Hauser, M. 1998, ApJ, in press, astro-ph/9812182.
\bibitem[]{1999ApJ...517..565P} Perlmutter, S., et al. 1999, \apj, 517, 565
\bibitem[]{1998MNRAS.298.1133P} Pozzetti, L. , Madau, P. , Zamorani, G. , Ferguson, H. C. \& Bruzual A., G. 1998, \mnras, 298, 1133
\bibitem[]{1996A&A...308L...5P} Puget, J.-L., Abergel, A., 
Bernard, J.-P., Boulanger, F., Burton, W.B., Desert, F.-X. \& Hartmann, D. 
1996, \aap, 308, L5 
\bibitem[]{renzini:a} Renzini, A. 1997, ApJ, 488, 35
\bibitem[]{1998AJ....116.1009R} Riess, A. G., et al. 1998, \aj, 116, 1009
\bibitem[]{1979Natur.281..635R} Rowan-Robinson, M., Negroponte, J. \& Silk, J. 1979, \nat, 281, 635
\bibitem[]{stog} Stognienko, R., Henning, Th. \& Ossenkopf, V. 1995, A\&A, 296, 797
\bibitem[]{vogeley97} Vogeley, M. 1997, ApJ, submitted, astro-ph/9711209
\bibitem[]{wick:a} Wickramasinghe, N., Edmunds, M., Chitre, S., Narlikar, J., \& Ramadurai, S. 1975, { Ap\&SS}, { 35}, L9
\bibitem[]{1996Ap&SS.240..157W} Wickramasinghe, N. C. \& Wallis, D. H. 1996, \apss, 240, 157
\bibitem[]{wollm:a} Wollman, E. 1992, ApJ, 392, 80
\bibitem[]{wright81} Wright, E. L. 1981, ApJ, 250, 1
\bibitem[]{wright:c} Wright, E. L. 1982, ApJ, 255, 401
\bibitem[]{wright87} Wright, E. L. 1987, ApJ, 320, 818

\end{thebibliography}
\end{document}